\documentclass{aa}

\usepackage{graphicx}

\begin{document}

\title{Semi-analytical homologous solutions of the gravo-magnetic contraction}

\titlerunning{Gravo-magnetic contraction}


\author{P. Hennebelle}

\offprints{P. Hennebelle, \email{patrick.hennebelle@ens.fr}}

\institute{ Laboratoire de radioastronomie millim{\'e}trique, UMR 8112 du
CNRS, {\'E}cole normale sup{\'e}rieure et Observatoire de Paris,
 24 rue Lhomond, 75231 Paris cedex 05,
France \and Department of Physics and Astronomy, Cardiff University, PO Box 913,
 5 The Parade, Cardiff CF24 3YB, Wales, UK
}

\abstract{ We propose an extension of  the
 semi-analytical solutions derived by Lin et al. (1965) describing
the two-dimensional homologous collapse of a self-gravitating rotating cloud
having uniform density and spheroidal shape, 
which includes magnetic field (with important restrictions)
  and thermal pressure.
 The evolution of the cloud is  reduced to 
three  time dependent ordinary equations  allowing to 
conduct a quick and preliminary investigation of the cloud dynamics 
during the precollapse phase, for a wide
range of parameters. 
We apply our model to the collapse of a  rotating and magnetized oblate 
and prolate isothermal core. Hydrodynamical numerical simulations are
performed and comparison with the semi-analytical solutions is discussed.
Under the assumption that all cores are similar, 
an apparent cloud axis ratio distribution is calculated from the
 sequence of successive evolutionary states for each of a large set of
initial conditions.
 The   comparison with the
observational distribution of the starless dense cores belonging
to the catalog of Jijina et al. (1999)
 shows a  good agreement for the rotating and initially prolate cores 
(aspect ratio $\simeq 0.5$) permeated by an helical magnetic field 
 ($\simeq 17-20 \mu$G for a density of $\simeq 10^4$ cm$^{-3}$). 
\keywords{Accretion, accretion disks -- Gravitation -- Hydrodynamics
-- Magnetohydrodynamics (MHD) -- ISM: clouds} 
}

\maketitle

\section{Introduction}

The stars form through the gravitational collapse of 
 dense cores located in the molecular clouds. These condensations are 
complex and involve gravity, hydrodynamics, magnetic  and
thermal processes. 

The core dynamics has been  investigated numerically by
several authors.
Although the numerical approach undoubtedly allows a deep
understanding of the cloud dynamics, it does not replace the
analytical studies, first because analytical work gives more explicit
descriptions  and second because the complexity of numerical methods 
 keeps increasing, 
  making  a full exploration of the space parameters  and 
  their use by non numerical experts difficult.

Analytically, most of the efforts have been dedicated to the
description of  the isothermal 
gravitational collapse 
(Larson 1969, Penston 1969a, 1969b, Hunter 1977, 
Shu 1977, Whitworth \& Summers 1985,
 Bouquet et al. 1985). These solutions
which assume self-similarity are very useful to get a physical hint 
on how the collapse proceeds and to give strong benchmarks to the
numerical studies. However in spite of its importance, 
 very few analytical tools are available
to describe the dynamics of the precollapsing or starless phase. 
The present paper is an attempt to fill this gap.

In this paper, we first present  new solutions of the gravo-magnetic
condensation. They are an extension of  solutions
obtained by Lin et al. (1965) for the gravitational part and
simultaneously by Aburihan et al. (2001) and by the author (Hennebelle 2001)
 for the MHD part. We then  use these solutions to study the phenomenology 
of the collapse  of oblate and prolate, isothermal, magnetized and non
magnetized clouds and compare the result with the evolution obtained
 with numerical simulations. 
The solutions allow to calculate the cloud shape
evolution.
With the assumption that all cores are similar, distributions of 
apparent axis ratio are predicted. 
 We then compare these distributions
 with the apparent axis ratio distribution of the starless dense cores
belonging to the catalog of Jijina et al. (1999).

Indeed  the geometry of the dense cores is an important
issue since
it depends on the physical processes that drive their evolution. Previous
 investigations by Myers et al. (1991) and Ryden  (1996) conclude that the 
observational data are more consistent with axisymmetric prolate cores
 than with axisymmetric oblate cores. 
The same conclusion is reached by the recent study of Curry (2002) with
a different sample of cores and by Hartmann (2002) who shows that the
major axis of the cores in the Taurus cloud is preferentially aligned
with the filament in which they are embedded.  
Two recent studies, (Jones et al. 2001
and Goodwin et al. 2002) analyse the observational apparent axis ratio 
distribution derived from the catalog of Jijina et al. (1999) and 
conclude that the core apparent axis ratio distribution
is compatible with  triaxal cores but being more nearly oblate 
than prolate.
From a theoretical point of view, the equilibrium solutions of a rotating
and magnetized cloud (Mouschovias 1976, Tomisaka et al. 1988) present an oblate
 shape except if there is a substantial 
 toroidal magnetic component   (Tomisaka 1991)
or if the structure of the poloidal magnetic field is such that it 
compresses the cloud (Stahler \& Curry 2001). 
Curry (2000)  also shows that two-dimensional solutions of the isothermal 
Lane-Emden equation present a prolate configurations.
On the other hand, 
 recent studies show (Nakamura et al. 1993, Fiege \& Pudritz 2000a, 2000b)
 that prolate cores can be formed from a magnetized  filament
(with poloidal and toroidal field) through a combination of the gravitational 
and the magnetic sausage instabilities. 
Fiege \& Pudritz  (2000c) also demonstrate that the axis ratio of the 
prolate magnetized cores is compatible with the observational apparent
axis ratio distribution. \\

In the second Section of the paper, we derive the solutions and reduce
the self-gravitating MHD equations to three time dependent ordinary equations.
We discuss their mathematical and physical properties, their
potential and intrinsic weaknesses.
In the third Section, we solve  these  ordinary 
equations numerically and describe the phenomenology 
of the collapse of  an isothermal dense core. 
We also compare the semi-analytical solutions with the results
of the hydrodynamical numerical simulations of an isothermal dense core.
The fourth Section is devoted
 to the comparison  between the theoretical and observational 
apparent axis ratio distributions. A discussion and summary are given 
in the fifth part.

\section{Homologous solutions of the gravo-magnetic condensation}
In this section we  look for  self-similar solutions of 
Eqs.~(\ref{consmat})-(\ref{eqmag3})  and reduce this system 
into a system of three 
ordinary differential equations of the time.

In most of the  classic studies of self-similar solutions  
 (Larson 1969, Penston 1969a, 1969b, Hunter 1977, 
Shu 1977, Whitworth \& Summers 1985,
 Bouquet et al. 1985), a special and simple time dependence of the 
fields is assumed leading to  ordinary equations 
of the self-similar variable. These solutions
are indeed related to the  invariance of the self-gravitating hydrodynamical
equations through the  dilatation groups (see e.g. Olver 1986 and Ferrara
\& Shchekinov 1996 for an introduction to the application of Lie groups
to differential equations). 

In this paper, on the contrary, we first assume a special and simple 
spatial dependence and put all the complexity in the time dependence of
 the fields. This technique has been successfully applied by Lin et al. (1965)
to predict the eccentricity evolution of a cold cloud and by 
Hennebelle \& P\'erault (2000) to study the thermo-magnetic condensation. 

\subsection{The equations}
We consider the perfect MHD equations of  self-gravitating 
gas with polytropic equation of state and assume an axisymmetric geometry. 
With the  usual notations, we have in cylindrical coordinates:

\begin{eqnarray}
\partial _t \rho + {1 \over \varpi } \partial _\varpi 
 \left(\varpi  \,  V_\varpi \, \rho \right)     
+ \partial _z \left(  V _z  \rho \right) =0 , 
\label{consmat}
\end{eqnarray}

\begin{eqnarray}
\nonumber
\rho \left( \partial _t V_\varpi + V_\varpi \partial _\varpi V_\varpi
+  V_z \partial _z V_\varpi 
 - { V_\theta ^2 \over \varpi } \right) =
   - \partial _\varpi P \\
+ \rho \partial _\varpi \Phi  -  { 1 \over \mu _0 \varpi }
B _\theta  \partial _\varpi (\varpi B _\theta)  +
{ 1 \over \mu _0 } B _z \left( \partial _z  B_\varpi
 - \partial _\varpi B_z  \right),
\label{consmom1}
\end{eqnarray}

\begin{eqnarray}
\nonumber
\rho \left( \partial _t V_\theta + V_\varpi \partial _\varpi V_\theta
+  V_z \partial _z V_\theta + {V _\theta V _\varpi \over \varpi} \right) = \\
  { 1 \over \mu _0 \varpi } B _\varpi  \partial _\varpi (\varpi B_\theta) + 
{ 1 \over \mu _0 } B _z  \partial _ z  B_ \theta,
\label{consmom2}
\end{eqnarray}

\begin{eqnarray}
\nonumber
\rho \left( \partial _t V_z + V_\varpi \partial _\varpi V_z
+  V_z \partial _z V_z  \right) =
   -  \partial _z P \\
+  \rho \partial _z \Phi  - 
 { 1 \over \mu _0  } B _\theta  \partial _z  B_\theta - 
 { 1 \over \mu _0 } B _\varpi 
\left( \partial _ z  B_\varpi
 -  \partial _\varpi  B_z  \right),
\label{consmom3}
\end{eqnarray}

\begin{eqnarray}
\partial _t (P \rho ^{-\gamma}) + V _\varpi \partial _\varpi (P \rho ^{-\gamma}) + V _z \partial _z (P \rho ^{-\gamma}) = 0 , 
\label{consener}
\end{eqnarray}

\begin{eqnarray}
{ 1 \over \varpi} \partial  _\varpi  (\varpi  \partial _\varpi \Phi) +
 \partial^2 _{z^2} \Phi  = - 4 \pi G \,  \rho,
\label{eqpoisson}
\end{eqnarray}

\begin{eqnarray}
 { 1 \over \varpi} \partial  _\varpi  (\varpi B _\varpi) + \partial _z B _z  = 0 , 
\label{eqdivb}
\end{eqnarray}

\begin{eqnarray}
\partial _t B _\varpi - 
 \partial _z (  V_\varpi B _z - V_ z B _\varpi  ) = 0 , 
\label{eqmag1}
\end{eqnarray}

\begin{eqnarray}
\partial _t B _\theta + 
 \partial _z (V_ z B_\theta - V_\theta B_z )
 - \partial _\varpi  (V_\theta B_\varpi - V_\varpi B_\theta)  = 0,
\label{eqmag2}
\end{eqnarray}

\begin{eqnarray}
\partial _t B _z +  {1 \over \varpi  }
 \partial _\varpi (\varpi (V_\varpi B_z - V_z B_\varpi  ) ) =0.
\label{eqmag3}
\end{eqnarray}

\subsection{Reduction to time-dependent ordinary differential equations}
Let us define the Jeans frequency:
\begin{eqnarray}
\Omega _J &=&  \sqrt{4 \pi G \rho _0 } , 
\label{def1}
\end{eqnarray}
and
\begin{eqnarray}
t = \tau / \Omega _J,
\label{def2}
\end{eqnarray}
where $\rho _0$ is the gas density.
We  consider the following fields:

\begin{eqnarray}
\nonumber
\rho(\tau,\varpi,z) &=& \rho _0 d(\tau)  , \\
\nonumber
P(\tau,\varpi,z) &=& \rho _0 \Omega _J ^2 \left( P _c (\tau) -  \, 
p _\varpi (\tau) \varpi ^2  -  \, p _z (\tau) z^2 \right), \\
\nonumber
\Phi(\tau,\varpi,z) &=& - \Omega _J ^2 d(\tau) ( A _\varpi(\tau)  \varpi ^2 + A _z(\tau) z ^2) , \\ 
\nonumber
V _\varpi(\tau,\varpi,z) &=& { \dot{a} (\tau) \over a (\tau) }  \Omega _J \; \varpi , \\
\nonumber
V _\theta(\tau,\varpi,z) &=& \dot{\theta} (\tau) \Omega _J \; \varpi  , \\
\label{champs}
V _z (\tau,\varpi,z) &=&  { \dot{c} (\tau) \over c (\tau) } \;  \Omega _J \; z , \\
\nonumber
B _\varpi(\tau,\varpi,z) &=&  \sqrt{\mu _0 \rho _0 } \; \Omega _J \;  h_\varpi(\tau) \; \varpi  ,\\
\nonumber
B _\theta(\tau,\varpi,z) &=& \sqrt{\mu _0 \rho _0 } \; \Omega _J \; h _\theta (\tau) \;  \varpi ,\\ 
\nonumber
B _z(\tau,\varpi,z) &=&  B _c(\tau) + 
\sqrt{\mu _0 \rho _0 } \; \Omega _J \;  h _z (\tau) \; z .\\
\nonumber
\end{eqnarray}
where the dotted values denotes derivative against $\tau$ and where
$d, \, P_c, \, p _\varpi, \, p_z  \, , A _\varpi, \, A_z \, , a, \, c, 
\, \dot{\theta} \, , h _\varpi, \, h _z, \, h _\theta$, and  $B _c$ are functions
of $\tau$.

The definitions stated by Eqs.~(\ref{champs}) lead with 
Eqs.~(\ref{consmat})-(\ref{eqmag3})
to the following  ordinary differential equations: \\

Continuity: 
\begin{equation}
\dot{d} (\tau) + \left( 2 { \dot{a} (\tau) \over a(\tau)} 
+ { \dot{c} (\tau) \over c(\tau)} \right) d(\tau) =0 ,
\label{eq1}
\end{equation}

Energy conservation:
\begin{equation}
{d \over d \tau} (P _ 0 (\tau)  d^{-\gamma} (\tau) ) = 0 ,
\label{eq2.0}
\end{equation}

\begin{equation}
{d \over d \tau} (p_\varpi (\tau) d^{-\gamma} (\tau)) +
 2 { \dot{a}(\tau) \over a(\tau) } (p _\varpi (\tau) d^{-\gamma} (\tau) ) =0 ,
\label{eq2}
\end{equation}

\begin{equation}
{d \over d \tau} (p_z (\tau) d^{-\gamma} (\tau) ) + 2 { \dot{c}(\tau) \over
c(\tau) } (p _z (\tau) d^{-\gamma} (\tau) ) =0 , 
\label{eq3}
\end{equation}

Poisson equation:
\begin{equation}
4 A _\varpi (\tau) + 2 A _z (\tau) = 1 ,
\label{eq4}
\end{equation}

Momentum conservation:
\begin{equation}
d(\tau)  \left( { \ddot{a} (\tau) \over a (\tau)} +
 \dot{\theta} ^2 (\tau) \right)   = 2 p_\varpi (\tau)  - 2 A _\varpi (\tau)
 d(\tau) ^2  - 2 h _\theta (\tau) ^2  ,
\label{eq5}
\end{equation}

\begin{equation}
d(\tau) \left( \ddot{\theta} (\tau) +  2 { \dot{a} (\tau) \over a(\tau)} 
  \dot{\theta} (\tau) \right)  = 2 h _\varpi (\tau) h_\theta(\tau) ,
\label{eq6}
\end{equation}

\begin{equation}
d(\tau)  {  \ddot{c} (\tau) \over c (\tau)}  =  2 p _z (\tau)
- 2 A _z (\tau) d(\tau) ^2 , 
\label{eq7}
\end{equation}

Nullity of magnetic divergence:
\begin{equation}
2 h _\varpi (\tau) + h _z (\tau) = 0 ,  
\label{eq8}
\end{equation}

Induction equation:
\begin{equation}
 \dot{h} _\varpi (\tau)  -  { \dot{a} (\tau) \over a(\tau)} h _z (\tau)  +
{ \dot{c} (\tau) \over c(\tau)}  h _\varpi (\tau) =0 ,
\label{eq9}
\end{equation}

\begin{equation}
\dot{h} _\theta (\tau) + \left(  2 { \dot{a} (\tau) \over a(\tau)}  +
 { \dot{c} (\tau) \over c(\tau)} \right) h _\theta( \tau) - 
 \dot{\theta}(\tau) (2 h _\varpi (\tau) + h_z(\tau) ) =0 ,
\label{eq10}
\end{equation}

\begin{equation}
\dot{h} _z (\tau) + 2 { \dot{a}(\tau) \over a(\tau) } h _z (\tau) -
 2 { \dot{c}(\tau) \over c(\tau) } h _ \varpi (\tau) = 0 ,   
\label{eq11}
\end{equation}

\begin{equation}
\dot{B} _c (\tau) +  2 { \dot{a} (\tau) \over a(\tau)}  B _c (\tau) =0.
\label{eq12}
\end{equation}

Eqs.~(\ref{eq1}),~(\ref{eq2.0}),~(\ref{eq2}),~(\ref{eq3}),~(\ref{eq9}),~(\ref{eq10}),~(\ref{eq11}) and~(\ref{eq12})
 are straightforwardly integrated
and lead to:
\begin{equation}
d(\tau) = {1 \over a(\tau) ^2 c (\tau) },
\label{dens}
\end{equation}

\begin{equation}
P _c (\tau) =  P ^0  d(\tau) ^\gamma , 
\label{press0}
\end{equation}

\begin{equation}
p _\varpi(\tau) = p ^0  _\varpi {d (\tau) ^\gamma  \over a (\tau) ^2  },
\label{press1}
\end{equation}

\begin{equation}
p _z (\tau) = p ^0 _z {d (\tau) ^\gamma  \over c(\tau) ^2  },
\label{press2}
\end{equation}

\begin{equation}
h _\varpi (\tau)=  { h ^0 _\varpi \over a(\tau) ^2 c (\tau) },
\label{magr}
\end{equation}

\begin{equation}
h _\theta (\tau)=   { h^0 _\theta \over a(\tau) ^2 c (\tau) },
\label{magthet}
\end{equation}

\begin{equation}
h _z (\tau)=  - 2 { h _\varpi ^0 \over a(\tau) ^2 c (\tau) },
\label{magz}
\end{equation}

\begin{equation}
B _c(\tau) = {B ^0 \over a(\tau) ^2  },
\label{magz}
\end{equation}
where $P^0$, $p _\varpi ^0$, $p _z ^0$, $h _\varpi ^0$, $h _\theta ^0$,
$B ^0$ are real numbers.

Defining $W(\tau) = a (\tau) ^ 2 \dot{\theta} (\tau)$, the
system of Eqs.~(\ref{eq1})-(\ref{eq12}) reduces
after some algebra,  to the three differential ordinary equations:

\begin{equation}
\ddot{a}(\tau)  =  
 { 2 p _\varpi ^0  \over a( \tau) ^ { 2\gamma -1} c(\tau) ^{\gamma -1} } 
+ { W ^2 (\tau) \over a(\tau) ^ 3 }
 - { 2 A _\varpi (\tau) \over a(\tau) c(\tau) }   
- {2 (h ^0 _\theta) ^2 \over a(\tau) c(\tau)} , 
\label{eqfinal1}
\end{equation}

\begin{equation}
\ddot{c}(\tau)  =  
 { 2 p _ z ^0  \over a( \tau) ^ { 2\gamma -2} c(\tau) ^{\gamma } } 
 - { 2 A _z (\tau) \over a(\tau)^2  },   
\label{eqfinal2}
\end{equation}

\begin{equation}
\dot{W} (\tau) = { 2 h^0 _\theta h^0 _\varpi \over  c (\tau) }. 
\label{eqfinal3}
\end{equation}

\subsection{Gravitational potential}
In the system of Eqs.~(\ref{eqfinal1})-(\ref{eqfinal3}), the gravitational
force (i.e. $A _\varpi$
and $A _z$) is unknown and must be specified.   
 Its value  indeed depends on  the boundary 
conditions. For simplicity, it can be assumed that $A _\varpi$
and $A _z$ remain constant along time, as it was the case in the 
studies carried out by Hennebelle (2001).
However the problem of boundary conditions has not been 
 addressed for these solutions. It is likely that
 because of boundary conditions, rotation
and magnetic field will induce 
an evolution of the shape of the cloud 
(and consequently a variation of $A _\varpi$
and $A _z$) for most 
of these solutions (if cylindrical geometry is assumed
 consistent solutions can be obtained, see Hennebelle 2003).

 More consistently, 
 if one considers elliptical boundary conditions,  i.e. a uniform axisymmetric 
spheroid, then one knows the gravitational potential as a function
 of the spheroid eccentricity $e$ (Mestel 1963, 
Lin et al. 1965, Chandrasekhar 1969, Binney \& Tremaine 1987).
 This allows to address properly 
 the problem of boundary conditions and to 
obtain self-consistent solutions. 

For an oblate spheroid, we have:
\begin{eqnarray}
\nonumber
A _\varpi (e)  &=&  {1 \over 4} { \sqrt{1 - e^2} \over e^2 } 
\left( { \rm{arcsin} (e) \over e} - \sqrt{1 - e^2} \right) , \\
A _z (e)  &=&  {1 \over 2} { \sqrt{1 - e^2} \over e^2 } 
\left(   {1 \over \sqrt{1 - e^2} } -  { \rm{arcsin} (e) \over e}  \right) ,
\label{oblate r}
\end{eqnarray}
and consequently:
\begin{eqnarray}
\nonumber
 e \rightarrow 0 &\Rightarrow& \left\{
\begin{array} {l}
A _\varpi \rightarrow 1/6 \\
A _ z \rightarrow 1/6
\end{array} \right. \\
e \rightarrow 1 &\Rightarrow& \left\{
\begin{array} {l}
A _\varpi \simeq (c /a)^2 \rightarrow 0 \\
A _z \rightarrow 1/2.
\end{array} \right.
\label{lim1}
\end{eqnarray}

For a prolate spheroid, we have:
\begin{eqnarray}
\nonumber
A _\varpi (e)  &=&  {1 \over 4} {1 - e^2 \over e^2 } 
\left(  {1 \over 1 - e ^2}  - {1 \over  2 e } 
 {\rm ln} \left( { 1 + e \over 1 - e} \right) \right) , \\
A _z (e)  &=&  {1 \over 2} {1 - e^2 \over e^2 } 
\left( {1 \over  2 e } 
 {\rm ln} \left( { 1 + e \over 1 - e}   \right) -1 \right). 
\label{prolate r}
\end{eqnarray}
and:
\begin{eqnarray}
\nonumber
 e \rightarrow 0 &\Rightarrow& \left\{
\begin{array} {l}
A _\varpi \rightarrow 1/6 \\
A _ z \rightarrow 1/6
\end{array} \right. \\
e \rightarrow 1 &\Rightarrow& \left\{
\begin{array} {l}
A _\varpi \rightarrow 1 / 4 \\
A _z  \simeq (a/c)^2 {\rm ln}(a/c)  \rightarrow 0
\end{array} \right.
\label{lim2}
\end{eqnarray}

 $l_\varpi$ and $l_z$ being the length of the 
radial and the polar axis respectively,  
the eccentricity is given by:
\begin{equation}
e(\tau) = \sqrt{ 1 - \left( {l_z \over l_\varpi} \right)^2 },
\end{equation}
for an oblate spheroid ($l_\varpi > l_z$), and by:
\begin{equation}
e(\tau) = \sqrt{ 1 - \left( {l_\varpi \over l_z} \right)^2 },
\end{equation}
for a prolate spheroid ($l_\varpi < l_z$).

The lengths of the axis evolve with time according to:
\begin{equation}
l_\varpi (\tau) =  l_\varpi(0)  a (\tau)  \; , \;
l_z (\tau) =  l_z(0)  c (\tau). 
\label{ellipse_ev}
\end{equation}

\subsection{Physical interpretations}
\label{discussion}
The system of Eqs.~(\ref{eqfinal1})-(\ref{eqfinal3}) is of 
second order in $a$ and $c$ and of  first order in $W$. 
It  thus depends on 5 
initial conditions, however one can assume without any restriction 
that $a(0)=1$ and $c(0)=1$. We thus have the 3 
initial conditions, $\dot{a}(0)$, $\dot{c}(0)$,
and $W(0)$  which respectively represent
the initial  radial velocity, axial velocity and rotation.

We also have the 5 parameters $P ^0, \gamma, 
 h^0 _\varpi$, $h^0 _\theta$ and $e(0)$ 
(as we will see with Eq.~(\ref{frontiere}), $e(0), \, P ^0, \, p ^0_\varpi$
and $p ^0 _z$ are indeed related)
which represent respectively 
the initial  thermal pressure, the polytropic index,
the initial magnetic poloidal field and magnetic toroidal field and the 
initial eccentricity.

The system of Eqs.~(\ref{eqfinal1}),~(\ref{eqfinal2}) and
(\ref{eqfinal3})  is much 
simpler than the system of Eqs.~(\ref{consmat})-(\ref{eqmag3})
and in spite of important physical restrictions, allows to study 
various non-linear aspects of the gravitational contraction.
Before solving Eqs.~(\ref{eqfinal1})-(\ref{eqfinal3}) numerically,
and in order to anticipate the numerical results, 
we discuss the physical meaning and mathematical behaviour of each 
of the terms in these three equations.

\subsubsection{The radial momentum conservation}
Eq.~(\ref{eqfinal1}) derives from the
 radial momentum equation, the first term of the 
right-hand side is the thermal pressure, the second  is the centrifugal
force, the third is
the gravitational force,  and the last is the toroidal magnetic pressure.
The poloidal magnetic pressure does not appear since it cancels out 
(the poloidal magnetic field is force free) as it was
already the case in the studies of Aburihan  et al. (2001) and Hennebelle (2001).

Since the density is uniform, the thermal pressure gradient 
is due to a non uniform 
temperature field. Although such temperature gradients are expected if
the gas is opaque  to its radiation,
   in the case of  dense cores that we will consider
in the next sections the temperature should be nearly constant. 
Rigorously speaking, these temperature gradients should be seen as ersatz
that allows to treat the thermal pressure of a uniform 
density cloud semi-analytically.
 



$\bullet$ If the collapse is spherical 
($p_\varpi ^0 = p_ z ^0$, $e(0)=0$, $W(0)=0$, $h^0 _\theta =0$), $a=c$
and Eq.~(\ref{eqfinal1}) can be integrated once  leading 
in the case  $\gamma \ne 1$ to:
\begin{equation}
{1 \over 2} \dot{a} ^ 2 = { 2 \over 3 - 3 \gamma } { p _\varpi ^0
\over a ^{3 \gamma - 3} } + {1 \over 3} {1 \over a } + K ,  
\label{eqdiscuss1}
\end{equation}
where $K$ is a constant which depends on the initial conditions.
If $\gamma \le 4/3$ the collapse is possible, since the gravitational force
 dominates the pressure force in the limit $a \rightarrow 0$. 
The collapse is not  possible if $\gamma > 4 /3$.

$\bullet$ If the cloud contracts  radially, we can
 assume that $c$ remains constant in the final stage.
 Remembering Eq.~(\ref{lim2}), Eq.~(\ref{eqfinal1}) can be further integrated 
(provided $\gamma \ne 1$) in the limit $a \ll c _0 $ ($c(\tau) \simeq c _0$):
  \begin{eqnarray}
\label{eqdiscuss2}
{1 \over 2} \dot{a} ^ 2 \simeq { 2 \over 2 - 2 \gamma } 
{ p _\varpi ^0 \over  a ^{2 \gamma - 2} c _0 ^{\gamma -1} } 
  &-&  { 2 W ^2(0)  \over a ^2} \\ 
 &-& \left( {1 \over 2 c_0} + {2 ({h^0 _\theta})^2  \over c_0 } \right) 
  { \rm ln } (a)  + K,
\nonumber
\end{eqnarray}
where $K$ again is an integration constant.
Therefore,  thermal pressure cannot support the 
cloud if $\gamma \le 1$
whereas if $\gamma > 1$, it  prevents the collapse. 
The centrifugal force  prevents the collapse  and the Lorentz force due
to the toroidal magnetic field adds up to  gravity.

$\bullet$ If the collapse occurs along the polar axis only 
($c \rightarrow 0$), 
 the pressure force is proportional to: $c ^{1 -\gamma}$, gravity 
is proportional to $c$
(remembering Eqs.~(\ref{lim1}), $A _\varpi \propto c ^2$)  and the 
toroidal force is proportional to $1/c$.  
Consequently, the radial component of the gravitational force
 becomes small compared to the thermal pressure whereas on the contrary,
 the toroidal magnetic force dominates the thermal pressure.

\subsubsection{The axial momentum conservation}
Eq.~(\ref{eqfinal2}) derives from the axial momentum equation.
It contains only the thermal pressure
 (first term of right-hand side) and gravity (second term)
since the magnetic forces in our geometry are  radial and azimuthal.

$\bullet$ If the cloud contracts radially ($a \rightarrow 0$), 
then with Eqs.~(\ref{lim2}), we find that the pressure force is proportional
to $a ^{2 - 2\gamma}$ and gravity to ${\rm ln}(a)$, consequently
 the thermal pressure dominates if $\gamma > 1$.

$\bullet$ If the collapse occurs along the pole ($c \rightarrow 0$),
assuming that $a$ remains constant ($a \simeq a _0$), 
we can integrate Eq.~(\ref{eqfinal2})  once
 (assuming that $\gamma \ne 1$):
  \begin{equation}
{1 \over 2} \dot{c} ^ 2 \simeq { 2 \over 1 -  \gamma } 
{ p _\varpi ^0 \over a _0 ^{2 \gamma -2} c ^{ \gamma - 1} } -
 {1 \over a _0 ^2} c + K,   
\label{eqdiscuss2}
\end{equation}
consequently, the  collapse into an infinitely thin disk is not
possible if $\gamma \ge 1$. As  pointed out in the conclusion
of Lin et al. (1965), at some point the cloud may rebound. 

\subsubsection{The azimuthal momentum conservation}
Eq.~(\ref{eqfinal3}) derives from the azimuthal momentum equation
(Eq.~\ref{consmom3}). The term of the right-hand side is the magnetic 
tension. If $h^0 _\varpi h ^0 _\theta \le 0$, it decreases the rotation
velocity (magnetic braking) and tends to increase it in the other case. 

\subsubsection{Boundary conditions}
In this section, we address the problem of the boundary conditions. \\

$\bullet$ Thermal pressure \\
The cloud boundary
 ($\varpi _b, z _b$) is
given by the condition that the thermal pressure vanishes. 
Eqs.~(\ref{champs}) lead to:
\begin{eqnarray}
P _c(\tau)  - p _\varpi \varpi _b ^2 - p _z z _b ^2 =0.
\end{eqnarray}
Thus, at the points
$(a(\tau) l_ \varpi (0), 0)$ and $(0, c(\tau) l _z (0) )$, we
 have with Eqs.~(\ref{press0})-(\ref{press2}):
\begin{eqnarray}
 P ^0 = p ^0 _\varpi  \left( { \varpi _b \over a } \right) ^2 
+   p ^0 _z   \left( { z _b \over c} \right) ^2 = p ^ 0 _\varpi l_\varpi(0) ^2
= p ^0 _z l_z(0) ^2.
\label{frontiere}
\end{eqnarray} 
Therefore the boundary is a spheroid having axis lengths equal to:
$ l_\varpi(0)=\sqrt{P ^0 / p ^0 _\varpi}$ an $ l_z(0)=\sqrt{P ^0 / p^0 _z}$.
We have consequently: $e(0) = \sqrt{1 - p^0 _ \varpi / p ^0 _z}$ for an oblate cloud
and $e(0) = \sqrt{1 - p^0 _z / p^0 _\varpi}$ for a prolate cloud.
In the case of a sphere, one has: $l_\varpi(0)=l_z(0) $ implying
$p^0 _\varpi = p^0 _z$. \\


$\bullet$ Magnetic field \\
The component of the magnetic field which is perpendicular to the
 cloud surface must be continuous through the surface 
whereas the component which is
tangent to the cloud surface can jump discontinuously from a value
to another leading to surface electric currents.

Since the density of the external medium is zero, the Lorentz force
must vanish in the external medium and the Alfv\'en speed is infinite.

This implies that the solutions given by Eqs.~(\ref{eqfinal1})-(\ref{eqfinal3})
with the condition stated by Eq.~(\ref{frontiere}) are compatible
with a poloidal magnetic field having the same structure 
(Eqs.~\ref{champs}) in the external medium and in the cloud 
($h_\varpi \ne 0$  implies that the magnetic field becomes infinite
when $r=\sqrt{\varpi ^2 + z^2} \rightarrow \infty$) and 
a vanishing toroidal magnetic component in the external medium.  
Note that the structure of the magnetic field does not 
necessarily 
follow the self-similar forms stated by Eqs.~(\ref{champs}) 
in the external medium and that other structures may be solution
of this problem.

The structure of the poloidal magnetic field in the external medium
 adjusts instantaneously as the boundary of the cloud moves,
 so that the field in the external medium
 remains force-free, since the Alfv\'en speed is infinite.

\subsubsection{Energy balance}
In the hydrodynamical case ($h^0 _\theta=h^0_\varpi=0$), the solution
is defined in a finite domain of space, and it is possible to find the
energy of the corresponding cloud. 

Let us multiply Eq.~(\ref{eqfinal1}) by $2 l_\varpi(0)^2 \, \dot{a}$, Eq.~(\ref{eqfinal2})
by $l_z(0) ^2 \dot{c}$ and add the two expressions.
Integrating the resulting equation, we obtain
when $\gamma \ne 1$:
\begin{eqnarray}
\nonumber
{1  \over 2} \left( 2 l_\varpi(0)^2 \dot{a} ^2 + l_z(0)^2 \dot{c} ^2 \right) +
{ 2 \over \gamma -1 } { P ^ 0 \over a ^{2 \gamma -2} c ^{\gamma -1}}
 + \\ { l_\varpi(0)^2 W^2(0) \over  a ^ 2} -  
 \left( 4 l_\varpi(0)^2 { A _\varpi \over c } + 2 l_z(0)^2 {A _z c \over 
 a ^2 } \right) = K,
  \label{energie}
\end{eqnarray}

For $\gamma = 1$, the integrated equation reads:
\begin{eqnarray}
\nonumber
{1  \over 2} \left( 2 l_\varpi(0)^2 \dot{a} ^2 + l_z(0)^2 \dot{c} ^2 \right) -
 2   P ^ 0  \ln ( a ^2 \,   c )
 + \\ { l_\varpi(0)^2 W^2(0) \over  a ^ 2} -  
 \left( 4 l_\varpi(0)^2 { A _\varpi \over c } + 2 l_z(0)^2 {A _z c \over 
 a ^2 } \right) = K,
  \label{energie_log}
\end{eqnarray}
which is a first integral of Eqs.~(\ref{eqfinal1})-(\ref{eqfinal2})
 and represents the total energy of the system. The first term is 
the kinetic energy, 
the second one is the thermal energy, the third one is the rotation energy and
the last one is the gravitational energy. 

If the magnetic field is non zero, the energy of the system is infinite
since the magnetic energy of the surrounding gas is infinite 
and it is not possible to find such a relation.

The  $\alpha$ and $\beta$ parameters, respectively the ratio
of the thermal to gravitational energy and of the rotational to gravitational
energy are equal to:

\begin{equation}
\alpha(\tau) = { 2 P ^0  \over \gamma -1  } { a ^{4 -2 \gamma } c ^{2- \gamma }
   \over ( 4 l_\varpi(0)^2  A _\varpi a ^2  + 2 l_z(0)^2 A _z c ^2  ) } ,
\label{alpha}
\end{equation}

\begin{equation}
\beta(\tau) = { l_\varpi(0)^2 W (0) ^2  c 
   \over ( 4 l_\varpi(0)^2  A _\varpi a ^2  + 2 l_z(0)^2 A _z c ^2  ) }.
\label{beta}
\end{equation}
In the case of an initially spherical cloud, we simply have: 
$\alpha(0)= 2 P ^0 /(\gamma-1)$ and $\beta(0)=W(0)^2$.

\subsection{Potential and weaknesses of the approach}
Although the derivation of Eqs.~(\ref{consmat})-(\ref{eqmag3}) into
Eqs.~(\ref{eqfinal1})-(\ref{eqfinal3}) is exact, 
the resulting reduction 
leaves aside some important physical effects. Nevertheless, it allows
to catch  some complex non-linear behaviours of the collapse
of a rotating and magnetized polytropic cloud at very little cost.
 This approach  complements  a 
full numerical simulation for which all the  parameter space cannot
be easily explored 
and which can possibly be affected by the numerical resolution and by
the numerical  scheme.
Finally such simple solutions  can be used 
to derive semi-analytic criteria (see e.g. Tsuribe \& Inutsuka 1999). 

We now discuss the potential and the weaknesses of this reduction.

\subsubsection{Potential}
Since the model is two dimensional, 
it is well suited to the study of the global  behaviour of 
a collapsing core including the radial and axial rebounds. 
It allows to study the effects of rotation on the collapse,
including the shape evolution and the subsequent variation of
 the gravitational potential. 

Since the equation of state is polytropic, 
the model also allows to study the influence of the 
polytropic index on the cloud dynamics.  

Since it includes magnetic field, it allows to catch some magnetic effects
as well, like toroidal pinching and magnetic braking.

\subsubsection{Weaknesses}
\label{weak}
The homologous reduction stated by Eqs.~(\ref{champs}) excludes any 
complex hydrodynamical features like shocks and turbulence. 
Due to  homology, the  dissipation by viscosity vanishes.

The temperature decreases from the center to  the cloud edge.
This prevents  the inwards propagation of the rarefaction wave that is usually
obtained in the numerical simulations of cloud collapse. 

The reduction stated by Eqs.~(\ref{champs}) leads
to a vanishing magnetic poloidal support. 
This clearly means that the solutions are not able to describe
the cloud evolution if the poloidal magnetic support is the
dominant process as it is the case for the collapse of a 
subcritical cloud controlled by ambipolar diffusion (e.g. Basu \& Mouschovias 
1994).

In these solutions the growth (or the decrease)
of the toroidal magnetic component due to the stretching of the 
poloidal component by differential (radial or axial) rotation 
vanishes, since the 
corresponding term cancels out in Eq.~(\ref{eq10}). This effect 
however is potentially
important since the collapse could be induced by the toroidal magnetic 
pressure (Habe et al. 1991, Tomisaka 1991). The axial support due to
the toroidal magnetic field vanishes as well.

\section{Isothermal cloud collapse: phenomenology}
\label{application}
In this section we investigate the dynamical behaviour of the solutions
of Eqs.~(\ref{consmat})-(\ref{eqmag3}) described by 
Eqs.~(\ref{eqfinal1})-(\ref{eqfinal3})  semi-analytically. 
The collapse of a  cold cloud (with no thermal pressure) has been studied
by Lin et al. (1965). They find that the oblate clouds collapse into
a disk whereas the prolate clouds collapse into a spindle. This is a 
purely gravitational effect. The gravitational force is higher along the
minor axis than along the major one and the initial asymmetry is amplified. 

\subsection{Application to dense cores}
As an application we will consider 
 the collapse of an isothermal prestellar cloud.
The density profiles of prestellar clouds have been found to be rather 
flat in the central part (e.g. Bacmann et al. 2000, Motte \& Andr\'e 2001)
having $\eta = d ln[\rho]/ dln[r] < 1$.
Consequently the uniform density approximation 
should be acceptable
to describe the dynamical evolution of their inner part during the 
precollapse phase.  Once the collapse phase begins, strong density
gradients develop, the present approach is no more valid and other
analytical techniques must be used to describe the collapse phase
(see e.g. the references quoted in the introduction).

 The typical physical parameters of the prestellar cloud are: 
temperature  $\simeq 10$ K, 
density $\simeq 10^4$ cm$^{-3}$ ($\rho \simeq 3 \, 10^{-20}$ g), 
 size $l _0 \simeq 0.1$ pc, 
adiabatic index $\gamma=1$, rotation ranging
from $\beta=0$ to $\beta=0.1$, where $\beta$ is the ratio of the rotational
energy to the gravitational energy.
The magnetic field unit is $\sqrt{\mu _0 \rho _0} \Omega _J l_0
\simeq$ 25 $\mu$G. 
The time unit, $1/\Omega _J$, is: $\simeq 2 \;  10^5$ years.

\subsubsection{Initial conditions}
As most of the observed cores are not spherical but have an observed aspect 
ratio of $\simeq 0.55$, 
we will consider initially oblate and prolate spheroids,
starting with an initial aspect ratio around 0.5.

For simplicity, we  start with a thermal pressure such that
the pressure force along the minor axis (along which it is stronger than along
 the major axis) is 3/4 of the gravitational force along the same axis.
The pressure along the major axis is then deduced from Eq.~(\ref{frontiere}).
We   start with initial velocity equal to zero, $\dot{a}(0)=0$
and $\dot{c}(0)=0$. 


 \subsubsection{Numerical simulations}
In order to assess the validity of the solutions and understand in
which regime they can be used, we perform
hydrodynamical numerical simulations using the SPH code described in 
Hennebelle et al. (2003). We use 50,000 particles to describe the
cloud and a constant external pressure equal to the initial 
 pressure of the cloud.
In the following, we simulate the evolution of clouds of
initially  uniform density. Note that this configuration is not
identical to the configuration described by
Eqs.~(\ref{eqfinal1})-(\ref{eqfinal3}). 
In the simulation the rarefaction wave rapidly leads to non uniform
density. Moreover the boundary pressure is not zero.

In order to have a realistic comparison, the mass of the cloud is chosen equal to 
one solar mass and the temperature to 10 K.

\subsection{Oblate cloud}
We first consider 
 non-rotating ($W(0)=0$), unmagnetized ($q(0)=h_\theta ^0 /a(0)^2/c(0) = 0$),
 oblate clouds and then clouds with various rotation speeds.

\subsubsection{Non-rotating cloud}

\begin{figure}
\includegraphics[width=8cm]{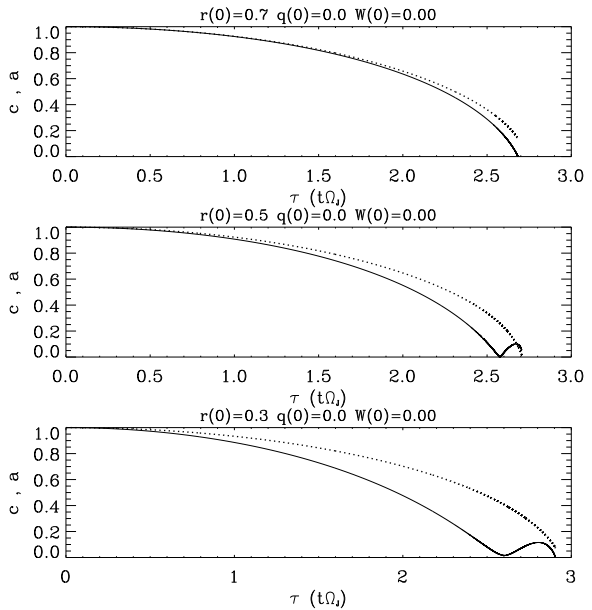}
\caption{Semi-analytical solutions 
of the collapse of an oblate isothermal ($\gamma=1$)
 cloud with initial axis ratio equal to 0.7 (upper panel),
0.5 (middle panel) and 0.3 (lower panel).
The evolution of the cloud radial
 axis contraction factor, $a(\tau)$ (dotted line),
 and the cloud polar axis contraction factor, $c(\tau)$ (full
line), are displayed. }
\label{obl}
\end{figure}

We consider  three different values of the  aspect ratio, namely 
$r(0) = l_z (0) / l_\varpi (0) =0.7, \; 0.5$ and 0.3.
The numerical integration is stopped when the density reaches the value
$10^5$. 
The results are  displayed in Fig.~\ref{obl}, 
 the evolution of $a(\tau)$, the radial axis (dotted line) and $c(\tau)$
the polar axis (full line) are presented. 

$\bullet$ In the  first case ($r(0)=0.7$), it is seen that, 
as in the study of Lin et al. (1965) gravity amplifies
the initial anisotropy and the cloud becomes more and more oblate. 
It eventually collapses into a disk at  $t \simeq 2.7 / \Omega _J$.

$\bullet$ In the second case, the minor axis collapses  more rapidly 
than the major axis 
 but after a strong contraction 
($c(0) / c _{\rm min} \simeq  10^3$), a rebound due to thermal pressure 
occurs at  $t \simeq 2.3 / \Omega _J$,
 the axial velocity becomes positive (outward motion) whereas the cloud
still contracts radially. 
Let us recall that in their conclusion, Lin et al. (1965) 
predict that the cloud
should oscillate if thermal pressure would be introduced.
 The cloud finally collapses into
a spindle at  $t=2.6 /\Omega _J$. 
Shortly before, the collapse along the polar axis
restarts. Note that the rebound is very stiff and  some care is required
 in the numerical resolution. 
We use adaptive time steps imposing that the maximum
variation of $a$ and $c$ between two steps is less than 1 \% and we
 checked that
the same result is obtained by requiring a variation of 0.5  \%. We also
check that the energy given by Eq.~(\ref{energie_log}) is well conserved.

$\bullet$ In the third case ($r(0)=0.3$), the rebound occurs still at
$t \simeq 2.6 / \Omega _J$ but the contraction is  weaker
 ($c(0) / c _{\rm min} \simeq  1/30$).
This is due to the fact that since the cloud is initially 
more oblate than in the previous case,  
the axial velocity is smaller  leading to a weaker 
contraction. Also the dynamical time scale along the minor axis is
shorter and the radial contraction is weaker when
the rebound occurs and the density is smaller by the time of this rebound.
After this rebound, the collapse
along the polar axis restarts and finally, the cloud collapses
into a disk, i.e. the density reaches our threshold, 
at $t \simeq 2.9 / \Omega _J$. 

\begin{figure}
\includegraphics[width=8cm]{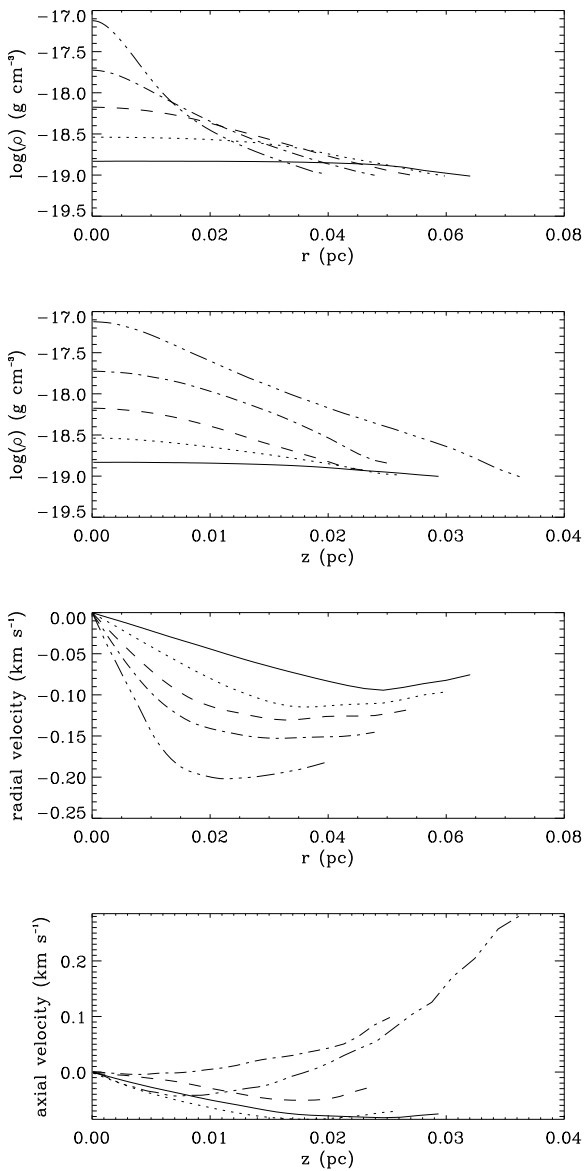}
\caption{Numerical simulation
 of a rotating oblate and isothermal ($\gamma=1$)
 cloud with initial axis ratio equal to 0.5. 
First panel displays the equatorial density, second panel
the density along the polar axis, third is the radial velocity at the
equator  and fourth is the axial velocity along the polar axis.
Full  line corresponds to t=0.89, 
dotted line to 1.34, dashed to 1.78, dot-dashed to 2.23 and
triple-dot-dashed to $2.67 /\Omega _J$.}
\label{obla_non_rot}
\end{figure}

$\bullet$ Fig.~\ref{obla_non_rot} presents the numerical simulation of 
a non rotating  oblate cloud (the initial aspect ratio equal is 0.5).
The initial length of the major axis is 0.07 pc. 
 Five time steps are displayed. It is seen that the velocity field 
 (both radial and axial components) can be well approximated by an 
homologous law in the inner part and that the inner part is also
reasonably flat during the precollapse phase. 
Because of the rarefaction wave, not
describe by the semi-analytical solutions, the density contrast 
in the outer part increases and the velocity is no more homologous. 
It is seen that at time $t=2.23 /\Omega _J$ the cloud bounces
axially because of the thermal pressure. Note that the inner part slows
down axially and stops collapsing 
($t=1.78$ and $2.23 / \Omega _J$) without  bouncing. 
The evolution of the radial component, $V _\varpi$ is monotonous in 
the inner and in the outer part.
At time  $2.67 / \Omega _J$ it is seen that the collapse
along the polar axis has restarted in the inner part whereas the
outer part is still bouncing.
Although this behaviour is too complex to be completely described by the 
semi-analytical solutions, they nevertheless undoubtedly reproduce
some complex features  of the cloud
evolution including the axial rebounds and subsequent collapse. 
Note that the largest density reached before the rebound  is 
significantly overestimated by the semi-analytical solution. This is
due to the thermal pressure which is not accurately
calculated. However, the duration of this phase of strong condensation
is short and does not alter much the subsequent evolution.

\subsubsection{Rotating oblate cloud}
\label{rot_obl}

\begin{figure}
\includegraphics[width=8cm]{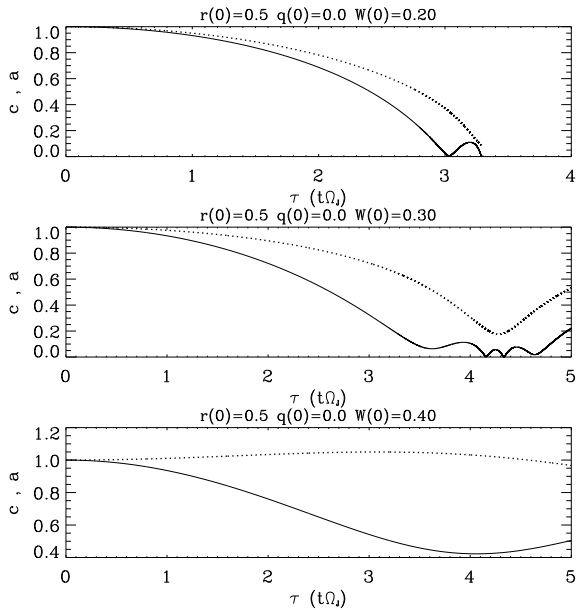}
\caption{Semi-analytical solutions 
of the collapse of a rotating oblate and isothermal ($\gamma=1$)
 cloud with initial axis ratio equal to 0.5.
 Upper panel is $W(0)=0.2$, middle one, $W(0)=0.3$ and 
 lower panel is $W(0)=0.4$.
The evolution of $a(\tau)$ (dotted line) and $c(\tau)$ (full
line) are displayed. }
\label{obl_rot}
\end{figure}

We consider the evolution of an initially oblate cloud with 
initial axis ratio $r(0)=0.5$ and investigate three initial
rotation values, $W(0)=0.2, 0.3$ and 0.4. The results are displayed
in Fig.~\ref{obl_rot}.

$\bullet$ The first case ($W(0)=0.2$) is similar to the evolution of the
non-rotating  clouds
 (see Fig.~\ref{obl}), except that due to the centrifugal support, 
the rebound occurs later ($t \simeq
3 / \Omega _J$).
The cloud collapses into a disk at  $t \simeq 3.3 / \Omega _J$.

\begin{figure}
\includegraphics[width=8cm]{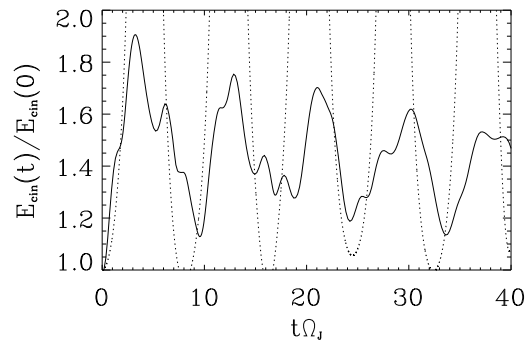}
\caption{The kinetical energy as a function of time. Full line is 
the numerical solution and dotted line is the semi-analytical one.}
\label{dissip}
\end{figure}

$\bullet$ In the second case ($W(0)=0.3$), it is found that the cloud does not collapse
but makes a series of rebounds. This is due to the fact that rotation 
strongly supports the cloud  and consequently the 
 contraction along the polar axis is too weak (density is lower and gravitational
force along the polar axis smaller than for the previous case)
 for allowing the density to reach
our  threshold. 
The cloud presents a quasi-periodic behaviour. 
Since Eqs.~(\ref{eqfinal1})-(\ref{eqfinal2}) are non-dissipative, the cloud
 never relax and never collapse. In a more realistic situation, the
number of rebounds  depend on the typical time scale of the dissipative 
processes. If  this time is short compared to the oscillation period, 
the rebounds will be damped rapidly (see Fig.~\ref{dissip} for a
numerical  estimate of this time).

$\bullet$ In the third case ($W(0)=0.4$), the centrifugal force is stronger
than for the previous case and the oscillations have a longer
period and a smaller amplitude.

\begin{figure}
\includegraphics[width=8cm]{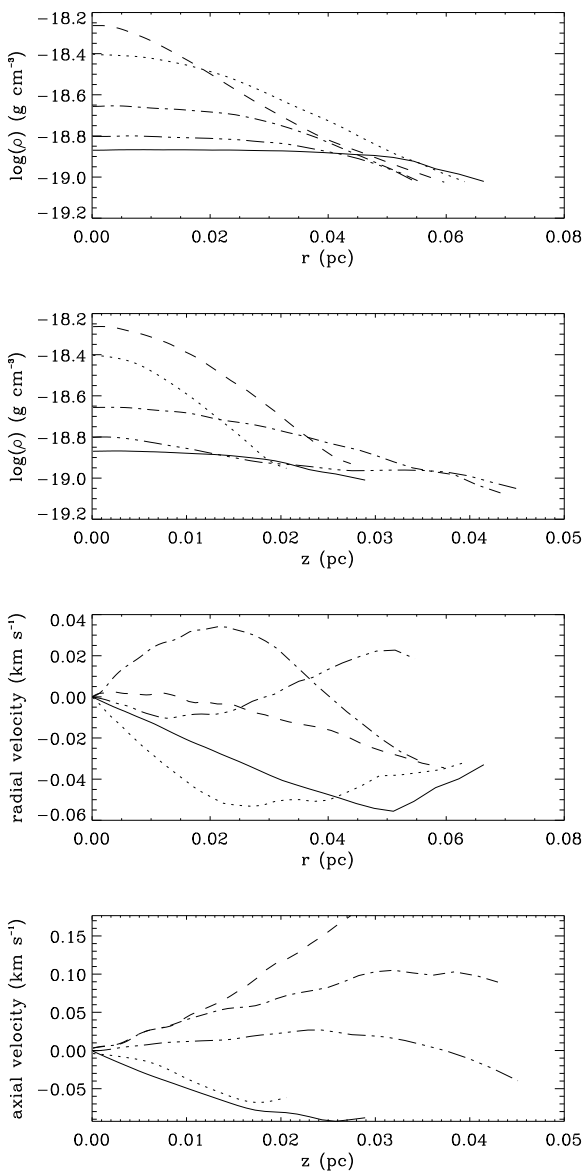}
\caption{Same as Fig.~\ref{obla_non_rot} for a rotating
oblate cloud (initial aspect ratio equal to 0.5 and $W(0)=0.3$).
Full  line corresponds to t=0.89, 
dotted line to 1.78, dashed to 2.68, dot-dashed to 3.57 and
triple-dot-dashed to $t=4.46 /\Omega _J$.}
\label{obla_rot}
\end{figure}

$\bullet$ Fig.~\ref{obla_rot} presents the numerical simulation of a
rotating ($W(0)=0.3$) initially oblate core ($r(0)=0.5$).
The   initial length of the major axis is 0.07 pc. 
As in the case studied in the previous section, it is seen that the 
homologous approximation is fair to describe the velocity field 
and that the cloud is rather flat in the inner part. 
The cloud starts to contract axially and radially. At time $t=2.68 
/ \Omega _J$ the cloud rebounds axially and at time $3.57 /\Omega _J$,
 the inner part
 rebounds radially whereas the outer part is still contracting. 
At time $t=4.46 /\Omega _J$ the outer part of the cloud expands
radially whereas the inner part is contracting. 
This behaviour is again very similar (though more  complex)
 to the case displayed in the
second panel of Fig.~\ref{obl_rot} where the axial rebound occurs at
time $\simeq 3.5 / \Omega _J$ and the radial one at $\simeq 4.2 /\Omega _J$.
The disagreement on the times is about 20 \%.

During the subsequent evolution, the cloud continue to bounce 
axially and radially several times and looses its energy slowly
as can be seen in Fig.~\ref{dissip} where the evolution of the
kinematical energy of the cloud is displayed. 
The minima correspond to the most expanded states
 and the maxima are close to the most condensed. It is seen that the 
global behaviour of the cloud is reproduced by the semi-analytical
solution during few oscillation cycles. The details however are  not 
accurately described. In particular, as already mentioned, the 
semi-analytical solutions overestimate the largest density reached
during the oscillation cycles. 

It is worth noting that such an oscillating behaviour may have been
recently observed in the dark cloud Barnard 68 by Lada et al. (2003)
where both systematic inwards and outward motions have been 
inferred.

\subsection{Prolate cloud}
We consider now the collapse of prolate clouds, i.e. clouds having 
filamentary shape with  an initial aspect ratio
 $r(0) = l_ \varpi(0) / l_z(0) = 0.5$. We  
first investigate the case of rotating and unmagnetized clouds
and then  rotating and magnetized filaments.

\subsubsection{Rotating unmagnetized cloud}
\label{rot_prol}

\begin{figure}
\includegraphics[width=8cm]{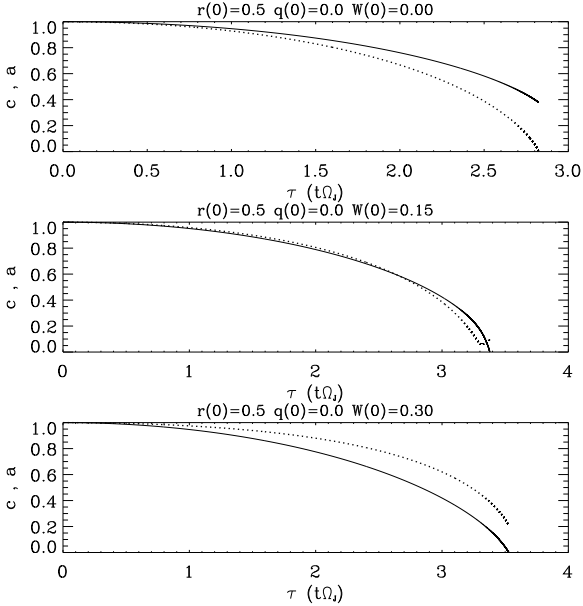}
\caption{Semi-analytical solutions 
of the collapse of a rotating prolate isothermal ($\gamma=1$)
 cloud with initial axis ratio equal to $r(0)=0.5$.
Upper panel is $W(0)=0$ (no rotation),
middle panel, $W(0)=0.15$ and lower panel, $W(0)=0.3$.
The evolution of $a(\tau)$ (dotted line) and $c(\tau)$ (full
line) are displayed. }
\label{prol_rot}
\end{figure}

Three values of the rotation speed are considered, $W(0)=0, \, 0.15$ and 
$0.3$. The results are displayed in Fig.~\ref{prol_rot}.

$\bullet$ The  first case (upper panel) is similar to the result of 
Lin et al. (1965).
The major axis collapses first since gravity is stronger in this
direction than along the major axis ($A _z / A _\varpi \simeq 0.42$)
 and the cloud  collapses into a spindle.

$\bullet$ In the second case ($W(0)=0.15$), it is seen that the collapse
of the minor axis is delayed by the  centrifugal support and that the
dynamics of the two axis is comparable.

$\bullet$ When the cloud rotates more rapidly ($W(0)=0.3$), the centrifugal
force is stronger and the cloud collapses more slowly in the
radial direction
 than along the polar axis. The cloud collapses into a disk at time
$t \simeq 3.5 / \Omega _J$.

\subsubsection{Rotating and magnetized cloud}
\label{rot_mag_prol}
\begin{figure}
\includegraphics[width=8cm]{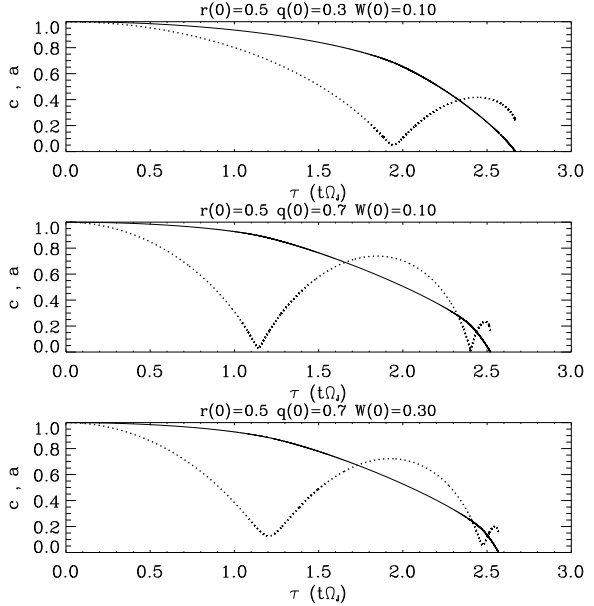}
\caption{Semi-analytical solutions 
of the collapse of a rotating and magnetized prolate isothermal ($\gamma=1$)
 cloud with initial axis ratio equal to $r(0)=0.5$.
Upper panel is $W(0)=0.1, q(0)=0.3$
middle panel, $W(0)=0.1, q(0)=0.7$ and lower panel, $W(0)=0.3, q(0)=0.7$.
The evolution of $a(\tau)$ (dotted line) and $c(\tau)$ (full
line) are displayed. }
\label{prol_rot_mag}
\end{figure}
We investigate now the collapse of a rotating and magnetized cloud.
For simplicity, we consider the same initial condition as in 
Sect.~\ref{rot_prol} and add a toroidal magnetic field with an
amplitude $q=0.3$ and 0.7.
The results are displayed in Fig.~\ref{prol_rot_mag}. 

$\bullet$ In the first case ($W(0)=0.1, q(0)=0.3$), the evolution is similar to
the non rotating prolate cloud ($W(0)=0, q(0)=0$) but the minor axis
 collapses more rapidly. This is due to the toroidal magnetic
field that  adds up to gravity in the equatorial plane but not along the
polar axis (initially we have, $A _z / (A _\varpi + q(0)^2) \simeq
0.29$). At time $t=1.9 / \Omega _J$ a radial rebound occurs because  rotation 
prevents the collapse.
 The cloud then becomes less and less prolate, 
until  $t \simeq 2.5 / \Omega _J$,
 where the collapse in the radial direction 
restarts. It eventually collapses into a disk at time $t \simeq
2.7 / \Omega _J$.

$\bullet$ If the toroidal magnetic field is stronger ($q(0)=0.7$),
the cloud is more strongly squeezed 
($A _z / (A _\varpi + q(0)^2) \simeq 0.12$) and the minor axis collapses
more rapidly than in the previous case and at  $t \simeq 1.2 / \Omega _J$, the 
first rebound occurs. By this time, the major axis has little evolved
because its dynamical time scale is slower.
At $t \simeq 1.9 / \Omega _J$, the collapse in the radial direction restarts
and the cloud makes a second rebound at time $t \simeq 2.4 / \Omega _J$.
It eventually  collapses into a disk at  $t \simeq 2.5 / \Omega _J$.

$\bullet$ The evolution of the more rapidly rotating cloud ($W(0)=0.3$) is 
very similar to the slowly rotating one. The only difference appears
when rotation becomes significant, i.e. during the contractions and the 
subsequent rebounds  at  $t \simeq 1.2$ and $2.5 / \Omega _J$. This is due to
the fact that in these two cases the centrifugal force is 
small compared to the toroidal pinching when the
cloud is weakly condensed. The centrifugal force
 plays an important r\^ole only when the 
cloud is significantly radially condensed.

Let us summarise our conclusions. 
The collapse of a prolate and magnetized core (with strong 
toroidal pinching) occurs slowly through gravity along the polar axis and rapidly 
through toroidal pinching and gravity in the radial direction. 
Since rotation prevents
radial collapse, few rebounds occur during the 
slow collapse along the polar axis and finally the cloud collapses into a disk.
Because of the toroidal magnetic field, the dynamical time scale
of the minor axis is much shorter than the  dynamical time scale of the
major one.
This process depends weakly on the rotation amplitude
 (as long as it is neither zero nor very large) and on the thermal pressure.


\section{Apparent shape of the clouds, comparison with observation}
In spite of the important restrictions listed in sect.~\ref{weak}, 
the solutions presented in this paper enable the study
of the cloud shape evolution during the precollapse phase. 
It is worth trying to 
compare the derived predictions with observations of dense cores
 traced by the NH$_3$ molecule (Jijina et al. 1999, Lee \& Myers 1999).
Since our simple model assumes a uniform density and does not include
complex processes like accretion or protostellar outflows, we 
select  the starless dense cores only from  the Jijina et al. (1999) catalog.

 Two recent studies (Jones et al. 2001, Goodwin et al. 2001) analyse the
observed apparent axis ratio distribution derived from this catalog
and conclude that this
distribution is compatible with dense cores being triaxal.

Since our model makes restrictive assumptions and since 
the Jijina et al. (1999) catalog
is not an homogeneous sample, the cores 
 having been observed with various instruments
and at various resolutions, the present analysis is rather tentative.
The idea is to see if a good fit can be obtained and  for which range 
of parameters. The result will not be more than indicative.

\subsection{Method}
The  first step of our analysis is to derive a shape
distribution from the time sequence of $a$ and $c$.
This is achieved   by counting the fraction of time
that one cloud spends with a given axis ratio. 

The second step is to project the axis ratio distribution along the 
line of sight.
Let $\psi$  be  the  axis ratio distribution,
and  $\phi$, the apparent (or observed) axis ratio distribution.  
The projection along the line of sight is  achieved according to the law
(Binney 1978, Fall \& Freck 1983):

\begin{equation}
\phi(r) = r \int _0 ^ r ds (1-s^2) ^{-1/2} (r^2-s^2) ^{-1/2} \psi(s) 
\label{proj_obl}
\end{equation} 
for an oblate distribution,

\begin{equation}
\phi(r) = r^{-2} \int _0 ^ r ds s ^2 (1-s^2) ^{-1/2} (r^2-s^2) ^{-1/2} \psi(s) 
\label{proj_prol}
\end{equation} 
for a prolate distribution.

The third step is to  calculate the $\chi ^2$ 
between this theoretical distribution
and the distribution  derived from the catalog of 
Jinina et al. (1999) which includes 79 starless cores.

We explore a large set of parameters:
\begin{description}
\item[-] the  ratio between the 2 axis,
 $r(0)=l_z(0) / l_\varpi(0)$
varies from 0.5 (oblate cloud) to 2.2 (prolate cloud),
 with a regular sampling of 0.1 leading to 18
values, 
\item[-] the magnetic toroidal intensity $q(0)$ varies 
from 0 to 1.5 (16 values), in physical units it ranges from
0 to $40 \mu$G.
\item[-] the rotation velocity, $W(0)$, from 0 to 0.6 
(7 values) which corresponds to $\beta$ ranging from 0 to 36\%. 
\item[-] the thermal pressure from $p _i ^0= ( A _i(0)) / 10$ to
$A _i(0)$ (10 values) where $i=z$ for oblate cores and
$i=\varpi$ for prolate cores.
\end{description}

Consequently, we solve $18 \times 16 \times 7 \times 10$ ($\simeq 2 \, 10^4$)
 times 
Eqs.~(\ref{eqfinal1})-(\ref{eqfinal2}) and calculate the corresponding apparent
axis ratio and the $\chi ^2$ with the observed distribution for each
of these solutions.

\subsection{Result}
\begin{figure}
\includegraphics[width=8cm]{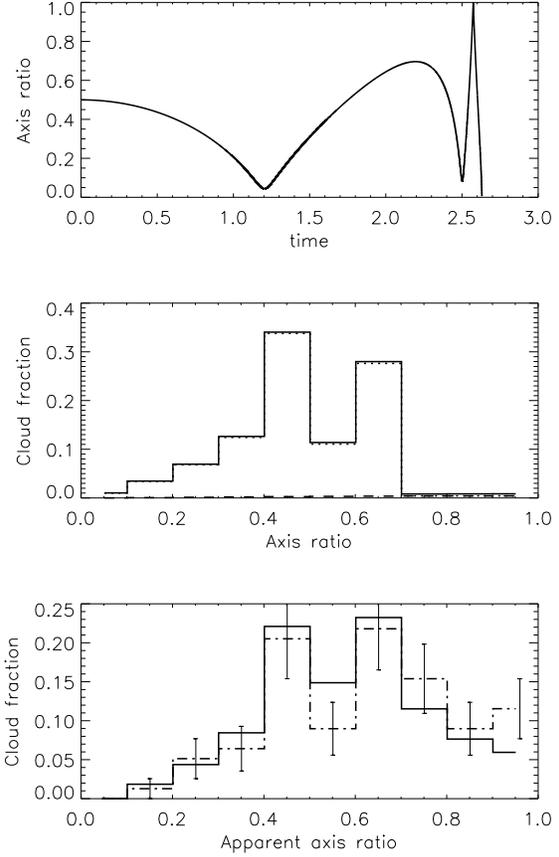}
\caption{Apparent axis ratio during the cloud evolution (upper panel),
axis ratio distribution (middle panel) and apparent axis ratio distribution
(lower panel). 
The dotted line (second panel) is the fraction of prolate core, the dashed
line, the fraction of oblate cores. In the third panel the dot-dashed 
curve is the apparent axis distribution of starless cores 
belonging to the catalog of Jijina et al. (1999) whereas the full line
is the theoretical distribution obtained for the
initial conditions  $l_\varpi(0)/l_z(0) = 0.5$, $q=0.7$ 
(magnetic toroidal intensity), $W(0)=0.3$ (rotation) and
 $p_\varpi ^0= A _\varpi(0)/7$ (thermal pressure). 
The value of $\chi ^2$ is 0.56.}
\label{stat_bo}
\end{figure}

The best agreement obtained is for 
$l_ \varpi (0)/l_ z(0) = 0.5$, 
$q=0.7$, $W(0)=0.3$, $P_c= A _\varpi(0)/7$, which is
very close to the second and third cases we have considered in 
Sect.~\ref{rot_mag_prol} (prolate magnetized cores).
The theoretical (full line) and observational (dot-dashed line) are
 displayed in the third panel of 
Fig.~\ref{stat_bo}. 
The agreement is  good and  better than in the study of 
Jones et al. (2001) and Goodwin et al. (2002).
The corresponding value of $\chi ^2$ is: 0.56 (assuming an 
 observational Poisson noise and no theoretical noise). 
The distribution depends
weakly on the value of rotation and thermal pressure and is much sensitive
to the initial axis ratio $r(0)=l_ \varpi (0) / l_ z(0)$ 
and to the magnetic toroidal intensity.

We find 179 cases with $\chi < 1$ (87 have $\chi<0.8$)
 similar to the best case. They have
$l_z(0)/l_\varpi(0)$ equal to 2 to 2.2, $q$ equal to 0.6 to 0.9
and all possible values of  $W(0)$ and $P_c$.

We also find 59 cases that belong to another  family.
They are oblate ($l_z(0)/l_\varpi(0)=0.5$) fast rotating cores 
($W(0) \ge 0.4$) similar to the cases presented in
Sect.~\ref{rot_obl}.
 However, the smallest $\chi$ is 0.85 and most of them 
have $\chi > 0.9$. The most important disagreement is due to the fact 
that too many clouds with  apparent axis ratio close to 1, are found.

\subsection{Discussion}
The  apparent axis ratio distribution derived from the evolution 
of the prolate magnetized filaments
 agrees  with the observed distribution.

 The  rotating  oblate cores  agree less well with
 the observational apparent axis ratio  distribution. 
Indeed the core remains oblate and Ryden (1996) shows
 that an oblate core distribution cannot explained the 
observed apparent axis ratio distribution.

We would like to emphasize the fact that the axis ratio distribution depends 
weakly on the initial 
rotation and  thermal pressure values (almost all  values of
these parameters are compatible with the observed axis ratio distribution)
 and strongly on the initial 
axis ratio and on the toroidal magnetic field. Consequently, the agreement
is obtained by adjusting two parameters only whereas there are 9 bins in
the fitted distribution. Therefore the fact that we obtain a parameter 
range that fits  the data is not a trivial consequence of   a large number
of parameters. Indeed there would not be acceptable fits without a
toroidal magnetic field.

 The most striking feature of  the unprojected axis ratio distribution 
of the  example displayed in  Fig.~\ref{stat_bo} is that it is bimodal
(with two peaks around 
$p_1 \simeq 0.5$ and $p _2 \simeq 0.7$).  
   This is due to the fact that the cloud spends 
 a long  time, with its initial aspect ratio and a long time with 
the aspect ratio it has after the first rebound.
This feature is smoothed by the projection effect and the apparent
axis ratio distribution is only slightly or marginally bimodal.

The data of Jijina et al. (1999) are compatible with (though marginally
statistically significant) a starless cores apparent axis 
ratio distribution being bimodal 
(see the dot-dashed curve of Fig.~\ref{stat_bo} or the starless cores
axis ratio distribution showed in Goodwin et al. 2002).
 The same trend appears also in two of the subsamples  
studied by Ryden (1996) and in the data studied recently
by Curry (2002) although in all of these samples the poor statistic 
  precludes a definite answer. More data are needed to confirm or to 
 reject this feature.




\section{Conclusion}
New solutions of the gravo-magnetic contraction have been presented. 
They describe the evolution of a uniform density cloud with spheroidal
shape and can be applied in various situations. These solutions allow to 
explore a wide set of parameters and constitute a useful complement to 
numerical simulations. They can also be used as benchmarks in problems
 involving gravitational collapse. 

We have explored the phenomenology of the gravitational contraction of
isothermal dense cores described by these semi-analytical solutions
 with special emphasis on the axial and  radial  rebounds. 
Numerical simulations have been performed to compare 
these solutions with the evolution 
of an isothermal initially uniform cloud and it has been found that
they  can reproduce  some aspects of 
cloud evolution  during the precollapse phase. The
most important disagreement is the cloud outer part which is
significantly influenced by the rarefaction wave not taken into account in the
semi-analytical solutions. 

For each of a large set of initial values, we have calculated the apparent
axis ratio distribution resulting from the evolution of the core 
and tentatively,
 compared it to the observational one derived 
from the catalog of Jijina et al. (1999).  We find that the
 axis ratio distribution 
of a collapsing rotating, prolate (initial aspect ratio $\simeq 0.5$)
 magnetized ($\simeq 17-20 \mu$G  for a density of $\simeq 10^4$
cm$^{-3}$)  filament is in  good agreement
with the observational distribution (see Fig.~\ref{stat_bo}). 
 Definite conclusion cannot be drawn at this stage 
however  since our model is very simplified
 and (mhd) numerical studies are required to confirm
the results. Also   the observational core sample is not
homogeneous and the statistical significance  remains poor.
Finally   alternative
distributions of evolving or static cores can  fit the observational 
data as well. 
Nevertheless, we believe that prolate cores permeated by an
 helical magnetic field deserve  further investigations.

\section{acknowledgement}
I thank Philippe Andr\'e,
  Simon Goodwin, Derek Ward-Thompson and Anthony Whitworth for
stimulating discussions. 
I acknowledge a critical reading of the manuscript by Michel
P\'erault.
I thank Jason Fiege, the referee for very insightful and detailed reports.
I gratefully acknowledge the support of an European Commission 
Research Training Network under the Fifth Framework Programme (No. 
HPRN-CT2000-00155) and a CNES fellowship.

\end{document}